\documentclass [12pt] {article}
\usepackage{a4}

\usepackage{amsfonts}
\usepackage{amsmath}

  \def\pp{{\mathchoice
          {
              \kern 1pt%
              \raise 1pt
              \vbox{\hrule width5pt height0.4pt depth0pt
                    \kern -2pt
                    \hbox{\kern 2.3pt
                          \vrule width0.4pt height6pt depth0pt
                          }
                    \kern -2pt
                    \hrule width5pt height0.4pt depth0pt}%
                    \kern 1pt
           }
            {
              \kern 1pt%
              \raise 1pt
              \vbox{\hrule width4.3pt height0.4pt depth0pt
                    \kern -1.8pt
                    \hbox{\kern 1.95pt
                          \vrule width0.4pt height5.4pt depth0pt
                          }
                    \kern -1.8pt
                    \hrule width4.3pt height0.4pt depth0pt}%
                    \kern 1pt
            }
            {
              \kern 0.5pt%
              \raise 1pt
              \vbox{\hrule width4.0pt height0.3pt depth0pt
                    \kern -1.9pt  
                    \hbox{\kern 1.85pt
                          \vrule width0.3pt height5.7pt depth0pt
                          }
                    \kern -1.9pt
                    \hrule width4.0pt height0.3pt depth0pt}%
                    \kern 0.5pt
            }
            {
              \kern 0.5pt%
              \raise 1pt
              \vbox{\hrule width3.6pt height0.3pt depth0pt
                    \kern -1.5pt
                    \hbox{\kern 1.65pt
                          \vrule width0.3pt height4.5pt depth0pt
                          }
                    \kern -1.5pt
                    \hrule width3.6pt height0.3pt depth0pt}%
                    \kern 0.5pt
            }
        }}

  \def\mm{{\mathchoice
                       {
                             \kern 1pt
               \raise 1pt    \vbox{\hrule width5pt height0.4pt depth0pt
                                  \kern 2pt
                                  \hrule width5pt height0.4pt depth0pt}
                             \kern 1pt}
                       {
                            \kern 1pt
               \raise 1pt \vbox{\hrule width4.3pt height0.4pt depth0pt
                                  \kern 1.8pt
                                  \hrule width4.3pt height0.4pt depth0pt}
                             \kern 1pt}
                       {
                            \kern 0.5pt
               \raise 1pt
                            \vbox{\hrule width4.0pt height0.3pt depth0pt
                                  \kern 1.9pt
                                  \hrule width4.0pt height0.3pt depth0pt}
                            \kern 1pt}
                       {
                           \kern 0.5pt
             \raise 1pt  \vbox{\hrule width3.6pt height0.3pt depth0pt
                                  \kern 1.5pt
                                  \hrule width3.6pt height0.3pt depth0pt}
                           \kern 0.5pt}
                       }}

\catcode`@=11
\def\un#1{\relax\ifmmode\@@underline#1\else
        $\@@underline{\hbox{#1}}$\relax\fi}
\catcode`@=12

\let\du=\du                     

\def\a{\alpha}

\def\f{\phi}

\def\p{\pi}

\def\O{\Omega}

\def\bo{{\raise-.5ex\hbox{\large$\Box$}}}               
\def\pa{\partial}                                       
\def\TH{{\raise.2ex\hbox{$\displaystyle \bigodot$}\mskip-4.7mu \llap H \;}}
\def\face{{\raise.2ex\hbox{$\displaystyle \bigodot$}\mskip-2.2mu \llap {$\ddot
        \smile$}}}                                      

\def\abs#1{\left| #1\right|}                    
\def\leftrightarrowfill{$\mathsurround=0pt \mathord\leftarrow \mkern-6mu
        \cleaders\hbox{$\mkern-2mu \mathord- \mkern-2mu$}\hfill
        \mkern-6mu \mathord\rightarrow$}
\def\dvec#1{\vbox{\ialign{##\crcr
        \leftrightarrowfill\crcr\noalign{\kern-1pt\nointerlineskip}
        $\hfil\displaystyle{#1}\hfil$\crcr}}}           

\def\frac#1#2{{\textstyle{#1\over\vphantom2\smash{\raise.20ex
        \hbox{$\scriptstyle{#2}$}}}}}                   
\def\sfrac#1#2{{\vphantom1\smash{\lower.5ex\hbox{\small$#1$}}\over
        \vphantom1\smash{\raise.4ex\hbox{\small$#2$}}}} 
\def\bfrac#1#2{{\vphantom1\smash{\lower.5ex\hbox{$#1$}}\over
        \vphantom1\smash{\raise.3ex\hbox{$#2$}}}}       
\def\afrac#1#2{{\vphantom1\smash{\lower.5ex\hbox{$#1$}}\over#2}}    

\def\[{\lfloor{\hskip 0.35pt}\!\!\!\lceil}
\def\]{\rfloor{\hskip 0.35pt}\!\!\!\rceil}

\def\du#1#2{_{#1}{}^{#2}}

\def\fracm#1#2{\hbox{\large{${\frac{{#1}}{{#2}}}$}}}

\def\un{\underline}
\def\fracmm#1#2{{{#1}\over{#2}}}

\def\low#1{{\raise -3pt\hbox{${\hskip 0.75pt}\!_{#1}$}}}

\newskip\humongous \humongous=0pt plus 1000pt minus 1000pt
\def\caja{\mathsurround=0pt}
\def\eqalign#1{\,\vcenter{\openup2\jot \caja
        \ialign{\strut \hfil$\displaystyle{##}$&$
        \displaystyle{{}##}$\hfil\crcr#1\crcr}}\,}
\newif\ifdtup

\def\pl#1#2#3{Phys.~Lett.~{\bf {#1}B} (19{#2}) #3}
\def\np#1#2#3{Nucl.~Phys.~{\bf B{#1}} (19{#2}) #3}
\def\prl#1#2#3{Phys.~Rev.~Lett.~{\bf #1} (19{#2}) #3}

\parskip=\medskipamount                 
\thicklines                         

\begin{document}
\thispagestyle{empty}

{\hbox to\hsize{
\vbox{\noindent KL~--~TH~ ~01/08   \hfill  February 2002 \\
hep-th/0109081                     \hfill  revised version }}}

\noindent
\vskip1.3cm
\begin{center}

{\Large\bf Engineering a Bosonic AdS/CFT Correspondence
\footnote{Supported in part by the `Deutsche Forschungsgemeinschaft'}}\\
\vglue.3in

Sergei V. Ketov, Thorsten Leonhardt and Werner R\"uhl

{\it Department of Theoretical Physics\\
     Erwin Schr\"odinger Stra\ss e \\
     University of Kaiserslautern}\\
{\it 67653 Kaiserslautern, Germany}
\vglue.1in
{\sl ketov,tleon,ruehl@physik.uni-kl.de}
\end{center}

\vglue.2in
\begin{center}
{\Large\bf Abstract}
\end{center}
\noindent
We search for a possible bosonic (i.e. non-supersymmetric) string/gauge theory 
correspondence by using ten-dimensional IIB and 0B strings as a guide. Our 
construction is based on the low-energy bosonic string effective action 
modified by an extra form flux.  
The closed string tachyon can be stabilyzed if the AdS 
scale $L$ does not exceed certain critical value, $L<L_{\rm c}$. We argue that
the extra form may be non-perturbatively generated as a soliton from 3-string 
junctions, similarly to the known non-perturbative 
(Jackiw-Rebbi-{}'tHooft-Hasenfratz) mechanism in gauge theories. The stable
$\textrm{AdS}_{13}\times S^{13}$ solution is found, which apparently
implies the existence of a 12-dimensional AdS-boundary conformal field
theory with the $SO(14)$ global symmetry in the large $N$ {}'t Hooft
limit. We also generalize the conjectured bosonic AdS/CFT duality to finite
temperature, and calculate the `glueball' masses from the dilaton wave
equation in the AdS black hole background, in various spacetime dimensions.

\newpage


\section{Introduction}

The (perturbatively defined, quantized) string theories can be either bosonic
or supersymmetric. The (open and closed) bosonic strings naturally live in
twenty-six dimensions, whereas the supersymmetric strings are ten-dimensional.
There is growing evidence that all five different supersymmetric strings are
related via dualities, while they also appear as certain limits of a single
supersymmetric (non-string) theory called M-Theory that is essentially
non-perturbative and has stable (BPS) branes \cite{mt}. Given the existence
of unique theory underlying all strings, there should exist a connection
between the supersymmetric M-theory and the bosonic strings too.
This raises a question whether the primordial theory must be supersymmetric,
and if not, one should then distinguish between the supersymmetric
(say, eleven-dimensional) M-theory
and the unified theory (U-Theory) of all strings and branes. In other words,
supersymmetry may not be fundamental but rather of dynamical origin in certain
compactifications of bosonic U-Theory. It should then be possible to
generate superstrings from bosonic strings.

In this paper we take this idea seriously and investigate some
of its implications from the viewpoint of a conjectured bosonic AdS/CFT 
correspondence.  The obvious objections against bosonic U-Theory are (i) the
apparent absence of fermions and (ii) the existence of a tachyon in the
bosonic string theory, either open or closed. To further justify our approach,
we first briefly address these fundamental issues.

The principal possibility of generating fermions from bosons in exactly
solvable two-dimensional quantum field theories (like sine-Gordon) is well
known since 1975 \cite{sin}.~\footnote{See, e.g.,
ref.~\cite{krev} for a recent review.}
This may also apply to the bosonic string in its world-sheet
formulation given by a two-dimensional conformal quantum field theory, as was
first noticed by Freund in 1984 \cite{fr}. Freund considered special
dimensional reductions of the bosonic string theory in 26 dimensions on
16-tori down to 10 uncompactified dimensions, and he argued that
the type-I open superstring theory with the $SO(32)$ gauge group
can be generated this way, with spacetime fermions originating as solitons via
the standard (Frenkel-Goddard-Olive) vertex operator construction \cite{fr}.
This implies the existence of a connection between the bosonic strings
and supersymmetric M-Theory since the latter is related to the type I
superstrings via the known chain of dualities \cite{mt}.

The world-sheet mechanism of generating spacetime fermions from bosons can be
complemented by the spacetime mechanism, which is known in the
(perturbatively) bosonic quantum gauge field theories with the $SU(2)$ gauge
group since 1976 due to Jackiw and Rebbi, Hasenfratz and {}'t Hooft
\cite{jrhh}. They argued that the bound state of a {}'t Hooft-Polyakov
monopole interacting with a free particle in a {\it half-integral}
representation of the unbroken $SU(2)_{\rm diag}$ symmetry is fermionic,
whereas it is bosonic in the case of an {\it integral}
representation.~\footnote{The {}'t Hooft-Polyakov monopole is well known to
preserve the diagonal subgroup
$SU(2)_{\rm diag}$ of \newline ${~~~~~}$
$SU(2)_{\rm space}\times SU(2)_{\rm isospin}$ \cite{krev}.}
Very recently, this field theory mechanism was generalized
to string theory by David, Minwalla and N\'u\~nez \cite{dmn}
who argued about the
non-perturbative appearance of fermions in various (perturbative) bosonic
string theories from 3-string junctions in higher spacetime
dimensions.

The significance of the {\it open} bosonic string tachyon was well understood
during the last two years after the crucial observation by Sen \cite{sen} who
first noticed that the open string vacuum can be viewed as the closed string
vacuum with an unstable D25-brane. It has then become evident that there
exists a stable minimum of the tachyon potential in the open bosonic string
theory at the value equal to minus the tension of that D25-brane. The tachyon
instability of the {\it closed}  bosonic string theory may be removed in the
strong coupling limit of this theory, as was recently argued by Horowitz and
Susskind \cite{hs} in the framework of hypothetical 27-dimensional {\it
Bosonic} M-Theory they proposed.

The crucial feature of all those M-Theory type constructions is the presence
of an extra background form flux. In the supersymmetric M-Theory the presence
of the 3-form is, of course, dictated by eleven-dimensional supersymmetry.
In Bosonic M-Theory, the existence of a three-form in 27 dimensions is
postulated \cite{hs}. The recently proposed non-supersymmetric type 0A string
theory in ten dimensions may also be interpreted as a type IIA string in the
background of a  Ramond-Ramond 2-form \cite{bg}. The tachyon instability of
the type 0A string apparently implies the tachyon decay into the stable
type IIA vacuum \cite{gs}.

In this paper we study the low-energy effective action of closed bosonic
strings in the background of a hypothetical gauge $(n-1)$ form $A$. The
perturbative spectrum of a closed bosonic string has only one such
(Kalb-Ramond) form with $n=3$. As was argued above, it is, however, possible 
that higher-$n$ antisymmetric tensor fields (including the self-dual ones)  
may be generated as solitons, similarly to the bound states involving 
instantons of $SU(N)$ gauge theories in even spacial dimensions or instantons 
of $SO(N)$ gauge theories in $(4n+1)$ spacial dimensions (that are especially
interesting for 26-dimensional bosonic strings), at least for sufficiently 
large $N$ ({\it cf.} ref.~\cite{dmn}). Since any field theory containing 
gravity is expected to be holographic, the bosonic AdS/CFT correspondence may 
apply at large $N$.
\vglue.2in


\section{Action and equations of motion}

Our starting point is the following field theory action in arbitrary even 
(Euclidean) spacetime dimensions $D$: 
\begin{align}
S & = \fracmm{1}{2 \kappa_0^2} \int d^D x \sqrt{\tilde g} \left\{
e^{-2\Phi} \bigl[ \tilde R - \fracmm{1}{12} \tilde H^2 + 4(\tilde \partial
\Phi)^2 \bigr] + \fracmm{1}{2 n!}\tilde F^2 \right\}~~.
\end{align}
Here $\tilde{g}_{MN}$ denotes a metric, $M=1,2,\ldots,D$, in the string frame,
$\tilde R$ is the associated Ricci scalar, $\Phi$ is a dilaton,
\begin{align}
(\tilde \partial \Phi)^2=\tilde g^{MN}\pa_M\Phi\pa_N\Phi~,
\end{align}
 and
\begin{align}
\begin{split}
\tilde F^2 = \tilde g^{M_1 N_1} \cdots \tilde g^{M_n N_n} F_{M_1 \ldots
M_n} F_{N_1 \ldots N_n}~~, \\
\tilde H^2 = \tilde g^{M_1 N_1} \cdots \tilde g^{M_3 N_3} H_{M_1 \ldots
M_3} H_{N_1 \ldots N_3}~~,
\end{split}
\end{align}
in terms of the field strength $F = d A$ of the gauge $(n-1)$ form $A$ and the
field strength $H = d B$ of the Kalb-Ramond two-form $B$.

The action (1) differs from the standard closed bosonic string effective 
action in the string frame \cite{Polch} by the last term mimicking the RR-type
 contributions in the type-IIB string theory. Thus, at the very bottom line,
we just study the field theory (1). We hope, however, that this field theory
may be connected to full bosonic string theory (see sect.~1).

The metric $g$ in the Einstein frame is related to the string frame metric
 $\tilde g$ via the Weyl transformation
\begin{align}
g_{M N} = e^{\frac{4}{D-2}(\Phi_0 - \Phi)} \tilde g_{M N}~~,
\end{align}
where $\Phi_0$ denotes the expectation value of the dilaton. We define
as usual $\phi := \Phi -\Phi_0$, so $\phi$ has the vanishing
expectation value. As is demonstrated in the next section, the
inclusion of a tachyon does not necessarily cause any damage, so we
can add the standard tachyonic action \cite{Polch} to our action in
the Einstein frame and thus obtain
\begin{align} \label{action}
\begin{split}
S_{\rm E} = & \fracmm{1}{2\kappa^2} \int d^D x \sqrt g \left\{  R -
\fracmm{1}{12} e^{-8\phi/(D-2)} H^2 - \fracm{4}{D-2} (\partial \phi)^2
-\fracmm{1}{2 n!} e^{\frac{2(D-2n)}{D-2}\phi} F^2  \right\} \\
 & - \fracmm{1}{2}  \int d^D x \sqrt g e^{-2\phi} \bigl[(\partial
T)^2 + m^2 T^2 \bigr] ~,
\end{split}
\end{align}
where $\kappa := \kappa_0 e^{\Phi_0}$, $m^2<0$ is the negative mass squared
of the tachyon $T$ and the contractions in $F^2$ and $H^2$ are performed with
the Einstein frame metric $g$.

The equations of motion derived from the action (5) are
\begin{itemize}
\item Einstein's equation,
\begin{align} \label{eom_metric}
\begin{split}
R_{M N} = & \frac{4}{D-2} \partial_M \phi \partial_N \phi + \kappa^2 e^{-2
\phi} \bigl[\partial_M T \partial_N T + \frac{m^2}{D-2} g_{M N} T^2 \bigr] +
\\ & + \frac{1}{2 n!}e^{\frac{2(D-2n)}{D-2}\phi} \left[ n F_M^{P_2
\ldots P_n} F_{N P_2\ldots P_n} - \frac{n-1}{D-2} g_{M N} F^2 \right] + \\
 & + \frac{1}{12} e^{-8 \phi /(D-2)} \bigl[ 3 H_M^{P_2 P_3} H_{N P_2 P_3} -
\frac{2}{D-2} g_{M N} H^2 \bigr]~,
\end{split}
\end{align}
\item Maxwell's equation for the $n-$form and the Kalb-Ramond field strength,
\begin{align} \label{eom_maxwell}
\partial_M \bigl(\sqrt g e^{\frac{2(D-2n)}{D-2}\phi} F^{M P_2 \ldots P_n}
\bigr) & = 0~ \textrm{and} \\
\partial_M \bigl(\sqrt g e^{-8 \phi/(D-2)} H^{M P_2 P_3} \bigr) & = 0~,
\end{align}
\item the tachyonic field equation,
\begin{align} \label{eom_tachyon}
\frac{1}{\sqrt g}e^{2\phi} \partial_M \bigl( \sqrt g e^{-2 \phi}
g^{MN} \partial_N T \bigr) - m^2 T =0~,
\end{align}
\item as well as the dilatonic field equation,
\begin{align} \label{eom_dilaton}
\Delta \phi + \frac{D-2}{4} \kappa^2 e^{-2 \phi} \bigl[(\partial T)^2 + m^2
T^2 \bigr] = \frac{D-2n}{8 n!} e^{\frac{2(D-2n)}{D-2}\phi} F^2 -
\frac{1}{12} e^{-8 \phi /(D-2)} H^2~,
\end{align}
\end{itemize}
where the Laplace-Beltrami operator $\Delta$, {\it viz.}
\begin{align}
\Delta \phi = \fracmm{1}{\sqrt{g}}\pa_M\left(\sqrt{g} g^{MN} \pa_N
\phi \right)~,
\end{align}
has been introduced.
\vglue.2in


\section{A stable non-supersymmetric solution}

We now observe that at $n = D/2$ (that is why we wanted an
evendimensional spacetime) the coupling between the dilaton $\phi$ and
the field strength $F$ in eq.~(\ref{eom_maxwell}) vanishes. The first
term on the right hand side of the dilaton equation
(\ref{eom_dilaton}) vanishes at $n = D/2$, too. The same situation
takes place in the type IIB supergravity in ten dimensions with the
self-dual RR five-form field strength, where dilaton also
decouples. We use the IIB superstring theory as our guide in the
nonsupersymmetric case (see also ref.~\cite{KleTsey} as regards the
similar case of type 0B strings in ten dimensions).

By using the standard (Freund-Rubin) compactification {\it Ansatz} \cite{dnp}
for our metric and the field strength $F$, we find that the $D$-dimensional
spacetime of the form $\textrm{AdS}_{D/2}\times S^{D/2}$ with the rest of the
fields set to zero is a solution to the equations of motion of Sect.~2,
namely,
\begin{align}\label{background}
\begin{split}
g & = g_{\textrm{AdS}_{D/2}} \oplus g_{S^{D/2}}~~, \\
\phi & = 0~, \\
T & = 0 ~, \\
H & = 0 ~, \\
F & = Q\, \textrm{vol}_{S^{D/2}}~~,
\end{split}
\end{align}
where vol$_{S^{D/2}}$ is the volume (top) form on the sphere $S^{D/2}$, and
$Q$  is a real constant.

The similar $\textrm{AdS}_5\times S^5$ solution (with a non-vanishing RR flux) 
\cite{ads} to the type-IIB supergravity equations of motion is believed to 
hold in the full type-IIB superstring theory, i.e. to all orders in $\a'$, 
being supported by unbroken supersymmetry \cite{adsr}. In our bosonic case 
the classical solution (12) may be modified after taking into account possible
 bosonic string corrections in higher orders with respect to $\a'$.

The stability of the field configuration (12) under small fluctuations,
\begin{align}\label{fluctuation}
\begin{split}
& g_{M N} \mapsto g_{M N} + h_{M N}~, \quad H \mapsto H + \delta H~,\quad F
\mapsto F + f~,
\\  & \qquad \qquad \qquad \phi \mapsto \phi + \delta \phi~,\quad T \mapsto T +
\delta T~, \end{split}
\end{align}
to the first order can be demonstrated as follows.

We first take a look at the right hand side of the Einstein equation
(\ref{eom_metric}) and observe that it is quadratic in the Kalb-Ramond
field strength as well as in the scalar fields and their derivatives. This
means that the variation of these terms to the first order about the
background (\ref{background}) vanishes. We are now left with merely the second
line as the right hand side of the Einstein equation (\ref{eom_metric}). We 
can now simply `borrow' a recent analysis of stability of Freund-Rubin 
compactifications in {\it non-dilatonic} gravity theories \cite{mit} where the
stability of the configuration (\ref{background}) for $g_{M N}$ and $F$ was
shown to the first order in the variations. Adding a Kalb-Ramond field, 
a dilaton and a tachyon does not change the geometry, at least in the given 
approximation.

In our case, the results of ref.~\cite{mit} are not enough because of
the additional scalars and the Kalb-Ramond field, which have to be
considered separately. We begin with the fluctuations of the
Kalb-Ramond field strength $H$. Inserting field variations
(\ref{fluctuation}) into the equation of motion (\ref{eom_maxwell})
together with the background (\ref{background}) results in the only
nonvanishing equation
\begin{align}
\partial_M \bigl( \sqrt g \delta H^{M P_2 P_3} \bigr) = \partial_M
\bigl( \sqrt g \partial^{[M} \delta B^{P_2 P_3]}\bigr),
\end{align}
or, in the language of differential forms,
\begin{align}
d \ast d \;(\delta B) = 0,
\end{align}
up to higher order terms in the fluctuations. In eq.~(15) we have introduced
$\delta B$ by $\delta H = d \delta B$, and we  have used the
fact that a variation $\delta$ and a derivative $d$ commute. Equation (15) is
just the equation of a free form field. Hence, by the same reasoning as
in ref.~\cite{mit}, we conclude that the background value $H=0$ is stable
under small fluctuations to the first order.

Varying the dilaton equation (\ref{eom_dilaton}) yields
\begin{multline}
(g-h)^{M N} \nabla_M \nabla_N (\phi + \delta \phi) + \\
+ \frac{D-2}{4} \kappa^2 e^{-2(\phi+ \delta \phi)} \bigl[ (g-h)^{M N}
\nabla_M (T + \delta T) \nabla_N (T + \delta T) +m^2 (T + \delta T)^2 \bigr]
 = \\ = -\frac{1}{12} e^{-\frac{8}{D-2}(\phi + \delta \phi)} (g-h)^{M_1
N_1} \cdots (g-h)^{M_3 N_3} (H + \delta H)_{M_1 M_2 M_3} (H +
\delta H)_{N_1 N_2 N_3}~.
\end{multline}
Being evaluated about the background (\ref{background}), this gives us a
fluctuation equation on $\delta \phi$ in the form
\begin{align}\label{fluc_dil}
\Delta (\delta \phi)  = 0~,
\end{align}
up to terms quadratic in the fluctuations. The Laplace operator in this
equation is supposed to be associated with the background metric of
eq.~(\ref{background}). Since this background is a product, the Laplace
operator in $D$ dimensions splits as
\begin{align}\label{split_lapl}
\Delta =\Delta\low{\textrm{AdS}_{D/2}} + \Delta\low{S^{D/2}}~.
\end{align}
The fluctuations can be expanded, as usual, in the spherical harmonics of
$S^{D/2}$ as follows:
\begin{align}\label{exp_sph}
\delta \phi = \sum_I \varphi^I(x) Y^I(y)~, & ~~~~{\rm where}&
x \in \textrm{AdS}_{D/2}~~& ~~{\rm and}& y \in S^{D/2}~~,
\end{align}
while the eigenvalue equation $\Delta_S Y^I(y) = - \lambda^I Y^I(y)$
with $\lambda \geq 0$ holds.
Inserting eqs.~(\ref{split_lapl}) and (\ref{exp_sph}) into eq.~(\ref{fluc_dil})
(and neglecting the higher order terms) yields
\begin{align} \label{analogy}
\Delta_{\textrm{AdS}} \varphi^I = \lambda^I \varphi^I~.
\end{align}
This implies the stability of the vanishing dilaton background under small
fluctuations to the first order, because of the non-negativity of all `masses'
$\lambda^I$. They all have to satisfy the {\it Breitenlohner-Freedman} (BF)
bound in the $\textrm{AdS}$ space \cite{bf} (see eq.~(\ref{BFbound}) below).

Finally, as regards tachyonic fluctuations, their computation due to the first
term in eq.~(\ref{eom_tachyon}) is quite similar to the dilaton case
considered above. In particular, the expansion of the tachyonic field
perturbation $\delta T$ in the spherical harmonics,
\begin{align}
\delta T = \sum_I t^I(x) Y^I(y)~,
\end{align}
has the same form as that of eq.~(\ref{exp_sph}). Hence, we have
\begin{align}
\Delta (\delta T ) = \sum_I \left[ \Delta_{\textrm{AdS}} t^I(x) - \lambda^I t^I(x)
\right]Y^I(y)~~.
\end{align}
The extra term linear in $T$ in eq.~(\ref{eom_tachyon}), upon evaluation about
the background configuration (\ref{background}) and dropping the higher
order terms, gives rise to a nonvanishing contribution from the mass term.
After taking all terms together, we find the following fluctuation equation
for a tachyon:
\begin{align}
0 & = \Delta (\delta T) - m^2 (\delta T) \nonumber \\
  & = \sum_I \left[ \Delta_{\textrm{AdS}} t^I(x) - \bigl( \lambda^I + m^2
\bigr) t^I(x) \right] Y^I(y)~.
\end{align}
To get a stable tachyon configuration, we have to satisfy the BF bound for all
modes $I$, i.e.
\begin{align}\label{BFbound}
\lambda^I  -\abs{m^2}  ~\geq~ m^2\low{\rm BF} = -\fracmm{1}{4 L^2}
\left(\fracmm{D}{2}-1\right)^2~~,
\end{align}
where we have explicitly indicated that the tachyonic mass parameter $m^2$ is
negative. Equation (\ref{BFbound}) apparently implies a restriction on the
scale parameter $L$ of the $\textrm{AdS}$ space, which is related to the
parameter $Q$ in the Freund-Rubin compactification {\it Ansatz}
(\ref{background}) for  the gauge field strength by
\begin{align}\label{VerhQzuL}
Q^2 = \fracmm{2(D-2)}{L^2}~~.
\end{align}

At this point we can study how the tachyon becomes dangerous when we make a
transition from $\textrm{AdS}_{D/2}\times S^{D/2}$ to flat spacetime, i.e.
towards the standard bosonic string theory,
in the limit $L \rightarrow \infty$.  Then the solution AdS$_{D/2}
\times S^{D/2}$ of Einstein's equation tends to $D-$dimensional
Euclidean spacetime. Because of eq.~(\ref{VerhQzuL}) the field strength $F$
vanishes in this limit. But what happens to the tachyon? The (fixed) tachyon
mass satisfies the BF bound until it is reached by increasing of $L$. Further
growth of $L$ gives rise to a violation of the BF bound by the tachyon mass,
which thus renders the whole spacetime unstable. We can thus consider
$L$ (or $Q$, respectively) as a moduli parameter of a family of the stable
AdS$\times S$ type solutions. We expect that the description of the theory in
terms of the action (\ref{action}) ceases to hold at the critical point where
the BF bound is reached. This apparently indicates on the existence of a phase
 transition at certain critical value $L_{\rm c}$ ({\it cf.} ref.~\cite{kle}).

We should also point out another interesting feature of field theory
in AdS space and its description in terms of the dual CFT. Consider
a scalar field $\Phi$ of mass squared $m^2$ in a $d+1$-dimensional
AdS spacetime. The BF bound arises from the formula
\begin{align}\label{Deltaofmsq}
\Delta = \fracmm{d}{2} \pm \sqrt{\fracmm{d^2}{4}+L^2 m^2}
\end{align}
that determines the parameter $\Delta$ in the propagator of
$\Phi$. This parameter should be a real number satisfying
\begin{align}
\Delta \geq \frac{d}{2}-1
\end{align}
in order to ensure normalizability of the scalar modes. Thus we see
that the BF bound is the necessary condition for the reality of
$\Delta$. Given $m^2 L^2>-(d/2)^2+1$, the normalizability condition
forces us to choose the plus sign in eq.~(\ref{Deltaofmsq}). Otherwise, both 
signs in eq.~(\ref{Deltaofmsq}) are allowed. As we approach the critical point
 $L_{\rm c}$, the two representations of the theory by the propagators with the
two possibilities for $\Delta$, respectively, join at $L_{\rm c}$.

Under the AdS/CFT duality \cite{ads,adsr} the parameter $\Delta$ translates 
into the conformal dimension of the dual field, while the normalizability 
condition goes into the positivity (unitarity) condition on the two-point 
functions of the dual CFT as a quantum field theory in the sense of Wightman 
\cite{HLMR}. In general, this is not enough to ensure positivity (unitarity) 
of the CFT in question. For example, a symmetric four-point function 
$\langle ABAB \rangle$ can be expanded into the conformal partial waves as 
\begin{align}
\langle ABAB \rangle \sim \sum_C (f^C_{AB})^2
\int dx dy \langle ABC(x) \rangle \langle 
C(x)C(y) \rangle^{-1}
\langle C(y)AB \rangle .
\end{align}
The conformal group representation of $C$ should be unitary while
the coupling constant squared $(f^C_{AB})^2$ should be positive. 

Summarizing above, a tachyon can be stabilized in the AdS$\times S$ type
background (with the other fields given above) provided the AdS scale does 
not exceed the certain value determined by the negative mass squared of the
tachyon. Unfortunately, this fact simultaneously restricts the applicability
of our result to the bosonic string theory because the low-energy bosonic 
string effective action is valid only if $\alpha'$ is small enough to suppress 
string loop corrections, while the spacetime curvature has to be small too, in
order to suppress strong gravity effects. The small curvature is needed to 
stabilize the tachyon in the sense of the BF bound, whereas the (negative) 
mass of the tachyon is proportional to $(\alpha')^{-1}$. The way out of this
difficulty may be fine tuning to get a domain, where $L$ is large enough while
$\alpha'$ is sufficiently small to let the low-energy effective action be 
valid while not spoiling the BF bound. This reasoning does not seem to be far 
fetched when comparing it with the type-0 string theory where a tachyon can be
stabilized for sufficiently small AdS radii in the dual CFT description
\cite{kle}.
\vglue.2in


\section{Bosonic AdS/CFT correspondence ?}

Having established the existence of a stable solution of the form
$\textrm{AdS}\low{D/2}\times S^{D/2}$, it is natural to speculate about the
existence of a bosonic AdS/CFT correspondence between gravity in the bulk
and a conformal field theory on the boundary, in a certain (with small 
spacetime curvature) limit. This correspondence may be just a sign of a more 
fundamental holographic duality between bosonic strings in the bulk and 
conformal field theory on the AdS boundary. The action of the AdS isometry
group on the AdS boundary is identified with the conformal group.

In the absense of supersymmetry  it is tempting to use type IIB superstrings
as a guide. As is usual in the AdS/CFT correspondence \cite{ads,adsr},
our stable solution $\textrm{AdS}\low{D/2}\times S^{D/2}$ should be identified
 with the near horizon limit of a black $p$-brane solution, with
$p=\left(\fracmm{D}{2}-2\right)$. Though this p-brane is not a BPS object 
since there is no supersymmetry to be partially broken, the cancellation of 
gravitational attraction and electromagnetic repulsion, needed for stability 
of the $p$-brane (see also eq. (\ref{magch1}) below), seems to be the natural 
substitute for the BPS property.

We are supposed to set $D=26$ in the case of bosonic strings,
which implies that our gauge field strength form $F$ should be a 13-form that
can also be self-dual in 26 dimensions, similarly to the analogous 5-form gauge
field strength of IIB superstrings. The boundary of the 13-dimensional AdS
space is 12-dimensional, so our considerations imply the existence of a
12-dimensional conformal field theory with a global $SO(14)$ symmetry, where
we have used the fact that $S^{13}=SO(14)/SO(13)$. It may not be accidental
that this 12-dimensional conformal gauge field theory is directly related to
F-Theory \cite{vafa} whose existence was motivated by the self-duality of
type IIB superstrings!

Imposing self-duality on $F$ in our non-supersymmetric setting may be
dangerous due to gravitational anomalies, since we have no mechanism
 at hand to cancel them. In the non-supersymmetric type 0B theory the
self-duality is needed in order to obtain conformal invariance of the
boundary theory up to two loop order \cite{KleTsey}. The gravitational
anomalies may cancel in the type 0B string theory, if the dual field theory is
an orbifold of $\mathcal{N} = 4$ supersymmetric Yang-Mills theory.

To establish a connection between the gauge coupling and the AdS radius $L$
in our bosonic case, we mimick the supersymmetric AdS/CFT correspondence
\cite{ads,adsr} with IIB superstrings in the bulk. Let us consider the magnetic
charge of $N$ D11-branes~\footnote{We define Dp-branes as the spacetime $(p+1)$
dimensional submanifolds where open bosonic \newline ${~~~~~}$ strings can end
({\it cf.} ref.~\cite{Polch}).}  stacked `on the top of each other'.
On the one hand, this charge is given by \cite{Polch}
\begin{align}\label{magch1}
g_{11} = N \tau_{11} \sqrt{16 \pi G_N}~,
\end{align}
where $\tau_{11}$ is the D11-brane tension and $G_N$ is Newton's constant.
They are related to `stringy' constants as follows \cite{Polch}:
\begin{align}\label{stringrel}
\begin{split}
\tau_{11} & = \fracmm{\sqrt \pi}{16 \kappa} (4 \pi^2
\alpha')^{(11-p)/2} = \fracmm{\sqrt \pi}{16 \kappa}~~,  \\
\kappa & = \sqrt{8 \pi G_N} = 2 \pi g_{\rm string}~~,
\end{split}
\end{align}
where $g_{\rm string}$ is the closed bosonic string coupling constant.
On the other hand,
the magnetic charge is just an integral of the field
strength $F$ over a sphere surrounding the D11-brane in the transverse
directions. The Freund-Rubin {¸\it Ansatz} (\ref{background}) now implies
\begin{align}\label{magch2}
g_{11} = \fracmm{1}{\sqrt{16 \pi G_N}} \int_{S^{13}} F =
\fracmm{Q}{\sqrt{16 \pi G_N}} \,  \Omega_{13}~~,
\end{align}
where $\Omega_{13}=2\pi^7/6!$ denotes the volume of $S^{13}$. Having
identified eqs.~(\ref{magch1}) and (\ref{magch2}), we use the
relations (\ref{stringrel}) to derive
\begin{align}\label{Qofgs}
Q  = \fracmm{N \tau_{11} 16 \pi G_N}{\Omega_{13}}
= \fracmm{90}{\pi^{11/2}} N g_{string}~~.
\end{align}

The Yang-Mills coupling constant of the gauge fields living in the worldvolume
of the D11-brane cluster is related to the D11-brane tension as
\begin{align}
g_{\rm YM}^2 = \tau_{11}^{-1}(2 \pi \alpha')^{-2}~.
\end{align}
This relation together with eq.~(\ref{stringrel}) implies a connection
between the Yang-Mills coupling constant and the string coupling constant in
the form
\begin{align}\label{gsofgYM}
g_{\rm string} = \frac{1}{8} \pi^{3/2} \alpha'{}^2 g_{\rm YM}^2~.
\end{align}

We recall that the extremal 11-brane solution is described by the
metric \cite{nbi}
\begin{align}
ds^2 = H^{-1/6} d{\vec x}{}^2 + H^{1/6} \bigl( dr^2 + r^2
d\Omega^2_{13} \bigr)~,
\end{align}
where we have introduced the notation $d{\vec x}{}^2 = dt^2 +
\sum_{j=1}^{11} (dx^j)^2$ and the metric on the to the 11-brane world
volume transverse space $dr^2 + r^2 d\Omega^2_{13}$. The `warp' factor
$H(r)$ becomes
\begin{align}
H(r) \equiv 1 + \fracmm{Q}{4 \sqrt 3 r^{12}} \stackrel{r \rightarrow 0}
{\longrightarrow} \fracmm{Q}{4 \sqrt 3 r^{12}} =
\fracmm{90}{32 \sqrt 3 \pi^4} N {\alpha'}{}^2 g_{\rm YM}^2 r^{-12}
\end{align}
in the near horizon limit $r \rightarrow 0$. In the last equality we have also
plugged in eqs.~(\ref{Qofgs}) and (\ref{gsofgYM}). We now introduce the string
length $l_{\rm string}^2 = \alpha'$, and obtain the following metric of the 
D11-brane in the near horizon approximation: 
\begin{align}\label{product}
ds^2 = \fracmm{r^2}{L^2} d{\vec x}{}^2 + \fracmm{L^2}{r^2} dr^2 +
L^2 d\Omega_{13}^2~,
\end{align}
where
\begin{align}
L^{12} = \fracmm{Q}{4 \sqrt 3 } = \fracmm{90}{32 \sqrt 3 \pi^4} N
g_{\rm YM}^2 l_s^4~.
\end{align}
Equation (\ref{product}) just describes $\textrm{AdS}_{13}(L) \times
S^{13}(L)$. In contrast to eq.~(\ref{VerhQzuL}), in the near horizon 
approximation, $L$ grows monotonically with $Q$.

We are now in a position to see the AdS/CFT correspondence at work in our
setting. Let's define the {}'t Hooft coupling,
\begin{align}
\lambda = g_{\rm YM}^2 N~,
\end{align}
and consider the large $N$ limit, $N \rightarrow \infty$ with $\lambda$ fixed,
which implies $g_{\rm YM}^2 \rightarrow 0$ and hence, $g_{\rm string}
\rightarrow 0$ too. This means that we can restrict ourselves to bosonic
string {\it trees}. To approach the strong {}'t Hooft coupling limit,
$\lambda \rightarrow \infty$, with a fixed length scale $L$, we must have
$l_{\rm string}^2=\alpha' \rightarrow 0$. This physically means that the
description of string theory in terms of particles, i.e. in terms of the
low-energy effective action (\ref{action}), becomes reliable. Thus we arrive
at the situation quite similar to the conventional AdS/CFT correspondence
in the supersymmetric type IIB context, but without supersymmetry.
\vglue.2in


\section{Glueball masses from higher dimensions}

It is rather straightforward to consider the thermodynamics of this 
12-dimensional conformal field theory by relating it to the near extremal
stable 11-brane. Perhaps even more exciting is the use of the conjectured
bosonic AdS/CFT duality at finite temperature, in order to get predictions for
the `glueball' masses as eigenvalues of the dilaton wave equation in the AdS
black-hole geometry near the horizon. This may shed light on a high
temperature expansion of the lattice QCD in higher (than four) spacetime
dimensions.

To calculate the $0^{++}$ glueball spectrum via Witten's method
\cite{wit}, we have to replace the $\textrm{AdS}$ factor of our
spacetime by an $\textrm{AdS}$ black hole of mass $M$. The metric of
such a $D/2-$dimensional black hole reads~\footnote{See, e.g.,
ref.~\cite{nbi} for details.}
\begin{equation}\label{metric}
ds_{\textrm{black hole}}^2 = V(r) dt^2 + \fracmm{1}{V(r)} dr^2 + r^2 d
\Omega_{n-1}^2~,
\end{equation}
where the function $V(r)$ is given by
\begin{equation}
V(r) = \fracmm{r^2}{L^2} + 1 - \fracmm{w_n M}{r^{n-2}}~~.
\end{equation}
We use the notation \cite{nbi}
\begin{equation}
w_n = \fracmm{16\p G_N}{(n-1)\O_{n-1}}~~,\quad{\rm and}\quad
 n=\fracmm{D}{2}-1~,
\end{equation}
where $G_N$ is the gravitational (Newton) coupling constant, and
$\O_{n-1}$ is the area of a unit sphere in $(n-1)$ dimensions.

Being a monotonic function, $V'(r) > 0$, it has exactly one zero (root) that 
we call a horizon $r_+$: $V(r_+)=0$.

The inverse (Bekenstein-Hawking) temperature of this black hole is \cite{nbi}
\begin{align}
\beta_0(r_+) = \fracmm{ 4 \pi r_+ L^2}{n r_+^2 + (n-2)L^2}~~.
\end{align}
In the large black hole mass limit, $M \rightarrow \infty$, the one in $V(r)$
can be dropped, so that we have
\begin{align}\label{Vofr}
V(r) \approx  \fracmm{r^2}{L^2} -\fracmm{w_n M}{r^{n-2}}~~.
\end{align}
After rescaling the coordinates as
\begin{align}
r = \biggl(\fracmm{w_n M}{L^{n-2}}\biggr)^{1/n} \,\rho \quad{\rm and}\quad
t = \biggl(\fracmm{w_n M}{L^{n-2}}\biggr)^{-1/n} \,\tau ~,
\end{align}
the black hole metric (\ref{metric}) takes the form
\begin{align}
ds_{\textrm{black hole}}^2 = \biggl(\fracmm{\rho^2}{L^2} -
 \fracmm{L^{n-2}}{\rho^{n-2}} \biggr) \, d
\tau^2 + \biggl(\fracmm{\rho^2}{L^2} - \fracmm{L^{n-2}}{\rho^{n-2}}
\biggr)^{-1}\,d \rho^2 + \biggl(\fracmm{w_n M}{L^{n-2}}\biggr)^{2/n} \,\rho^2
\,d \Omega_{n-1}^2~~. \end{align}
We now observe that the sphere in the large mass limit tends to the $(n-1)$
dimensional Euclidean space multiplied by a factor $\rho^2$. Further
rescaling of the coordinates as $\tau \mapsto L^2 \tau$ and $x_j \mapsto L 
x_j$ gives us the black hole metric
\begin{align} \label{blackback}
\fracmm{1}{L^2} \, ds_{\textrm{black hole}}^2 = \biggl( \rho^2 -
\fracmm{L^n}{\rho^{n-2}} \biggr) \, d \tau^2 + \biggl( \rho^2
-\fracmm{L^n}{\rho^{n-2}} \biggr)^{-1} \,d \rho^2 +\rho^2 \sum_{j=1}^{n-1}
dx_j^2~~.
\end{align}

Having replaced the AdS factor in $\textrm{AdS}_{D/2}\times S^{D/2}$ by the
black hole metric, we arrive at the full metric in the form
\begin{align}
\fracmm{1}{L^2} \, ds_{\textrm{full}}^2 = \biggl( \rho^2 -
\fracmm{L^n}{\rho^{n-2}} \biggr) \, d \tau^2 + \biggl( \rho^2
-\fracmm{L^n}{\rho^{n-2}} \biggr)^{-1} \,d \rho^2 +\rho^2 \sum_{j=1}^{n-1}
dx_j^2 + d \Omega_{n+1}^2~~.
\end{align}

The AdS black hole configuration can also be obtained from non-extremal branes
in the near horizon limit. To see this, let's consider the non-extremal 
$(D/2-2)$-brane configuration in our $D$-dimensional setting  
($n=D/2-1$, as above),
\begin{equation}
\eqalign{
ds^2 = & f(r) H^{-2/n}(r) dt^2 + H^{-2/n}(r) \sum_{j=1}^{n-1} (dx^j)^2
\cr & + f^{-1}(r) H^{2/n}(r) dr^2 + H^{2/n}(r) r^2 d\Omega^2_{n+1}
~,\cr}
\end{equation}
with $H(r) = 1 + (h/r)^{n} \rightarrow h^{n}/r^{n}$ for $r
\rightarrow 0$ and $f(r)=1-(r_0/r)^n$, where $r_0$ and $h$ are real
parameters, $r_0=0$ corresponds to the extremal situation, while $h$
is related to the radius of the forthcoming AdS-type space. We find
\begin{equation}
\eqalign{
ds^2  & =  f(r) \Bigl(\fracmm{h}{r} \Bigr)^{-2} dt^2 +
\Bigl(\fracmm{h}{r}\Bigr)^{-2} \sum_{j=1}^{n-1} (dx^j)^2 + f^{-1}(r)
\Bigl(\fracmm{h}{r} \Bigr)^2 dr^2 + h^2 d\Omega_{n+1}^2 \cr
& = V(r) dt^2 + V^{-1}(r) dr^2 + \fracmm{r^2}{h^2} \sum_{j=1}^{n-1}
(dx^j)^2 + h^2 d\Omega_{n+1}^2~,\cr} 
\end{equation}
where 
\begin{align}
V(r) = \fracmm{r^2}{h^2} - \fracmm{r_0^n}{h^2 r^{n-2}}~~ .
\end{align}
This just gives us an $(n+1)$-dimensional AdS black hole multiplied by an 
$(n+1)$-sphere after identifying $w_n M$ with $r_0^n/h^2$ and $h$
with $L$ ({\it cf.} our eqs. (\ref{metric}) and (\ref{Vofr})).

The classical equation of motion for the dilaton $\f$ in this black hole
background is similar to eq.~(\ref{analogy}),
\begin{equation} \label{masseq}
\Delta_{\rm black~hole} \varphi = \lambda^I \varphi~,
\end{equation}
though this time with the metric (\ref{blackback}). We only consider the s-wave
mode on the sphere, so that $\lambda^I=0$. We now look for solutions to
eq.~(\ref{masseq}), which are square integrable over the black hole spacetime
and correspond to a fixed momentum in the boundary theory. We further
demand that the solutions of interest are independent of $\tau$. This gives
rise to the following equation on $\varphi$~:
\begin{align} \label{masseq2}
\Delta_{\rm black~hole} \varphi = \partial_{\rho} \biggr( \left[
\rho^{n+1} - \rho L^n \right] \partial_{\rho} \varphi \biggr) + \rho^{n-3}
\sum_{j=1}^{n-1} \partial_j^2 \varphi = 0~.
\end{align}
The standard (`hedgehog') {\it Ansatz} for solutions to this equation is given
 by
\begin{align}\label{ansatz}
\varphi (\rho, \vec x) = f(\rho)\exp \left(i \vec k \cdot \vec x\right)~,
\end{align}
where $\vec x = (x_1,\ldots,x_{n-1})$ and $\vec k$ is a definite momentum in
 the boundary theory with $\vec k^2 = -m^2$ (of Euclidean signature!).
Inserting the {\it Ansatz} (\ref{ansatz}) into eq.~(\ref{masseq2}) and using
the fact that $\sum \partial_j^2 \varphi = -\vec k^2\varphi$, we find the
 ordinary second-order differential equation
\begin{align}\label{ode}
\rho^4 \left[ 1- \Bigl( \fracmm{L}{\rho} \Bigr)^n \right] f''(\rho) + \rho^3
 \left[ n+1- \Bigl( \fracmm{L}{\rho} \Bigr)^n \right] f'(\rho) + m^2 f(\rho) =
0~~,
\end{align}
where the primes denote differentiation with respect to the argument
$\rho$.

It is convenient to introduce the dimensionless variable $x=\rho/L$
and the dimensionless mass parameter $\tilde m = m/L$ \footnote{This
looks a little bit odd, but if one follows the derivation of eq.~(51) with
all the rescalings \newline ${~~~~~}$ done, one can check that $k_i$ and,
therefore, $m$ all have dimension of $L$ indeed.} in units of the AdS scale
$L$,
as well as define $y(x)=f(\rho)$. The equation (\ref{ode}) now takes the form
\begin{align} \label{Diffgl}
x^4 \left[ 1- \Bigl( \fracmm{1}{x} \Bigr)^n \right] y''(x) + x^3
\left[ n+1- \Bigl( \fracmm{1}{x} \Bigr)^n \right] y'(x) + \tilde m^2 y(x) =
0~~.
\end{align}

Our goal is a computation of those values of $\tilde m^2$ that lead to
square integrable and regular (at the horizon, near $x=1$)
solutions. Since we are unable to solve eq.~(\ref{Diffgl}) in a closed
form, we determine the relevant solutions numerically,~\footnote{We
used the standard software {\it Maple VI}~ for calculations on
computer.} in the form of power series defined inside the convergence
domains around the singular points of the differential equation
(\ref{Diffgl}) \cite{adsr}. The solutions given by the power series
must coincide in the overlaps of the convergence domains \cite{zy}. To
this end we observe that eq.~(\ref{Diffgl}) has singular points at
$x=0, \infty$ and the $n^{\textrm{th}}$ roots of unity. Therefore, we
only have three real (physical) singular points at $x \geq 0$. We do
not consider the root $x=0$, since we expect a spacetime singularity
there. Hence, we are left with the two singular points, $x=1$ and
$x=\infty$.

\subsection{The singular point $x=\infty$}

Changing the variables as $\xi = 1/x$ and $\eta (\xi) = y(x)$
transforms eq.~(\ref{Diffgl}) to the form
\begin{align}\label{Diffgl_inf}
\bigl( 1-\xi^n \bigr) \eta''(\xi) + \xi ^{-1} \bigl( 1-n-\xi^n \bigr)
\eta'(\xi) + \tilde m^2 \eta(\xi) = 0~.
\end{align}
The characteristic equation associated with eq.~(\ref{Diffgl_inf}) at
$\xi=0$ is given by
\begin{align}
\nu (\nu-1) + (1-n)\nu = 0~,
\end{align}
and it has two solutions, $\nu = 0$ and $\nu = n$. We only need $\nu =
n$ since the corresponding solution to eq.~(\ref{Diffgl}) decays for
$x \rightarrow \infty$ like $x^{-n}$ and is, therefore, normalizable,
whereas the other solution corresponding to $\nu =0$ is not. By using
the {\it Ansatz}
\begin{align}
\eta(\xi) = \xi^n \sum_{\mu \geq 0} b_{\mu} \xi^{\mu}~,
\end{align}
we get the following recursion relations for the coefficients
$b_{\mu}$:
\begin{align}
b_\mu = \fracmm{1}{\mu (\mu+n)} \Bigl( \mu^2 b_{\mu-n} - \tilde m^2 b_{\mu-2}
\Bigr)~.
\end{align}
With the initial value $b_0=1$ we thus obtain the whole power series
for $\eta$, which converges inside the circle $|\xi| < 1$, or
equivalently, $|x|>1$.

\subsection{The singular point $x=1$}

The characteristic equation associated with eq.~(\ref{Diffgl}) at
$x=1$ reads
\begin{align}
\nu(\nu-1)n + \nu n = 0~,
\end{align}
and it has a double root at $\nu = 0$. Hence, eq.~(\ref{Diffgl}) possesses
a pure power series solution,
\begin{align}\label{powers}
y(x) = \sum_{\mu \geq 0} c_{\mu} (x-1)^{\mu}~.
\end{align}
The second independent solution must have a logarithmic contribution,
$\log(x-1)$, which is not normalizable in the vicinity of $1$.
 Plugging the power series (\ref{powers}) into eq.~(\ref{Diffgl}),
we arrive at the recursive relations for the coefficients $c_{\mu}$,
\begin{multline}
c_{\mu} = - \fracmm{1}{n \mu^2} \sum_{j=1}^n \Biggl\{ \binom{n}{j} (\mu-j)
(\mu-j+n) + \tilde m^2 \binom{n-3}{j-1} \\
+ \binom{n}{j+1} (\mu-j)
(\mu-j-1) \Biggr\} c_{\mu-j}~.
\end{multline}
The convergence radius of the power series defined by these coefficients
equals the distance to the next singular point of the differential equation,
i.e. the minimum of $\{ 1,2\sin(\pi/n)\}$.

\subsection{Mass eigenvalues of the dilaton wave equation}

We now compare the two power series in the overlap region of their
convergence domains, in order to determine those values of the
parameter $\tilde m^2$, for which the two power series calculated
above define linearly dependent functions. In equivalent terms, we
look for zeros of the {\it Wronskian} as a function of $\tilde m^2$,
\begin{align}\label{wr}
{\rm Wronskian}(\tilde m^2) =\left. \det
\begin{bmatrix} \sum_{\mu \geq 0} c_{\mu} (x-1)^{\mu}
 & \sum_{\mu \geq 0} b_{\mu} x^{-(n+\mu)} \\
\frac{\partial}{\partial x} \bigl(\sum_{\mu \geq 0} c_{\mu} (x-1)^{\mu} \bigr)
& \frac{\partial}{\partial x} \bigl(\sum_{\mu \geq 0} b_{\mu} x^{-(n+\mu)}
\bigr) \end{bmatrix}  \right|_{x=x_0}~,
\end{align}
where the dependence of the right hand side upon $\tilde m^2$ is encoded in
the coefficients $c_{\mu}$ and $b_{\mu}$. The reference point $x_0$ is
supposed to belong to the overlap of the convergence domains. According to the
general theory of ordinary differential equations, the exact zeros of
Wronskian are independent upon a choice of the reference point $x_0$.

Though we cannot solve the recurrence relations analytically to all
orders, we can compute any given number of the coefficients on
the computer and truncate the power series at some finite number
$\mu_{max}$. Of course, because of the truncation, the positions of
the zeros of the Wronskian (\ref{wr}) are then only approximately
independent of the reference point $x_0$. For example, when taking
$x_0$ near the boundary of a convergence domain, the approximation is
becoming bad and the zeros of the Wronskian are getting shifted. Given
a sufficiently high cutoff $\mu_{max}$, the approximation of the power
series by finite polynomials appears to be good enough for our
purposes. We just plug these finite polynomials into the Wronsky
determinant (\ref{wr}), with $x_0$ being sufficiently far away from
the boundary of the convergence domains, and then we numerically
determine the zeros $\tilde m^2$. The results of our computation of
the first four zeros for various dimensions $n$ are summarized in
Table \ref{Tabell}.

We also verified that the zeros calculated in Table 1 are stable
against sufficiently small variations of the reference point $x_0$ and
the cutoff $\mu_{max}$.

\newpage

\begin{table}[htb]
\caption[Tabelle]{Glueball masses from higher dimensions}\label{Tabell}
\begin{center}
\begin{tabular}{|c||c|c|c|c|} \hline
n & $1^{st}$ zero & $2^{nd}$ zero & $3^{rd}$ zero & $4^{th}$  zero \\
\hline \hline
2 & 4.15 & 16.21 & 37.08 & 65.17 \\ \hline
3 & 7.41 & 24.96 & 52.56 & 90.21 \\ \hline
4 & 11.58 & 34.54 & 68.98 & 114.91 \\ \hline
5 & 16.49 & 44.73 & 85.54 & 138.92 \\ \hline
6 & 22.10 & 55.59 & 102.45 & 162.70 \\ \hline
7 & 28.37 & 67.09 & 119.80 & 185.54 \\ \hline
8 & 35.31 & 79.26 & 137.67 & 210.61 \\ \hline
9 & 42.90 & 92.09 & 156.09 & 235.03 \\ \hline
10 & 51.12 & 105.57 & 175.09 & 259.89 \\ \hline
11 & 59.97 & 119.71 & 194.69 & 285.22 \\ \hline
12 & 69.43 & 134.50 & 214.90 & 311.09 \\ \hline
13 & 79.52 & 149.93 & 235.74 & 337.50 \\ \hline
\end{tabular}
\end{center}
\end{table}

Our results for $n=4$ coincide with those of refs.~\cite{adsr,zy}.
\vglue.2in

\section*{Acknowledgements}

One of the authors (SVK) would like to thank the Department of Physics of the
University of Maryland in College Park, USA, 
 the Institute of Theoretical
Physics in Hannover, Germany, and the CIT-USC Center for Theoretical Physics
in Los Angeles, USA, for hospitality extended to him during different stages 
of this work.

\end{document}